\documentclass[a4paper,10pt]{article}

\usepackage{lscape}
\usepackage{cancel}
\usepackage{bm}
\usepackage{bbm}
\usepackage{multirow} 
\usepackage{amsfonts}
\usepackage{amssymb}
\usepackage{graphicx}
\usepackage{amsmath}
\usepackage[usenames,dvipsnames]{color}
\usepackage{hyperref}
\usepackage{epstopdf,epsfig}
\usepackage[left=1in,right=1in,top=1in,bottom=1in]{geometry}             % For good copy.

\hypersetup{
    colorlinks=true,         % false: boxed links; true: colored links, false is default
    linkcolor=blue,          % color of internal links, red is default
    citecolor=red,        % color of links to bibliography, 'green' is default
    urlcolor=Violet             % color of external links, cyan is default
}

\newtheorem{defi}{Definition}

\newtheorem{prop}{Proposition}

\def\be{\begin{equation}}
\def\ee{\end{equation}}
\def\bea{\begin{eqnarray}}
\def\eea{\end{eqnarray}}

\title{Semi-classical locality for the non-relativistic path integral in configuration space}
\author{\bf Henrique Gomes\footnote{\href{mailto:gomes.ha@gmail.com}{gomes.ha@gmail.com}}\\\it Perimeter Institute for Theoretical Physics\\ \it 31 Caroline Street, ON, N2L 2Y5, Canada}
\begin{document}
\maketitle
\begin{abstract}
In an accompanying paper \cite{path_deco}, we have put forward an interpretation of quantum mechanics based on a non-relativistic, Lagrangian 3+1 formalism of a closed Universe $M$, existing on timeless configuration space $\mathcal{M}$. However,  not much was said there about the role of locality, which was not assumed. This paper is an attempt to fill that gap. To deal with the challenges gauge symmetries may pose to a good definition of locality, I start by  demanding symmetries to have an action on $\mathcal{M}$ so that the quotient wrt the symmetries respects certain factorizations of $\mathcal{M}$. These factorizations are algebraic splits of $\mathcal{M}$ into sub-spaces $\mathcal{M}=\bigoplus_i \mathcal{M}_{O_i}$-- each factor corresponding to a physical sub-region $O_i$. This deals with kinematic locality, but locality in full can only emerge dynamically, and is not postulated. I describe conditions under which it can be said to have emerged. The dynamics of $O$ is independent of its complement, $M-O$,  if the projection of extremal curves on $M$ onto the space of extremal curves intrinsic to $O$  is a surjective map.  This roughly corresponds to $e^{i\hat{H}t}\circ \mathsf{pr}_{\mbox{\tiny O}}= \mathsf{pr}_{\mbox{\tiny O}}\circ e^{i\hat{H}t}$, where $\mathsf{pr}_{\mbox{\tiny O}}:\mathcal{M}\rightarrow \mathcal{M}_O^{\partial O}$ is a linear projection. This criterion for locality can be made approximate -- an impossible feat had it  been already postulated -- and it can be applied for theories which do not have hyperbolic equations of motion, and/or no fixed causal structure. 
When two regions are \emph{mutually independent} according to the criterion proposed here,  the semi-classical path integral kernel factorizes, showing cluster decomposition which is what one would like to obtain from any definition of locality.  
\end{abstract}

\section{Introduction}

In the previous paper \cite{path_deco}, I have introduced an interpretation of quantum mechanics based on path integrals in timeless configuration space. As my ultimate aim was to describe quantum cosmology, it was important for my purposes that the framework be adapted for the description of  a closed Universe, and in particular, of a non-relativistic gravitational system. 

However, to make \cite{path_deco} tractable and self-contained, I contented myself with discussing  consequences of the framework for foundational questions, such as describing a consistent picture of decoherence, records, coarse-graining, and the meaning and content of Born's rule, leaving aside issues of locality,  entanglement and the description of particular gravitational models amenable to that type of description. Since we most definitely experience subsystems, this was a glaring omission. The purpose of this accompanying piece is to fill some of these gaps. 

I intend to accomplish this by studying the dynamics in configuration space $\mathcal{M}$ -- in principle acting non-locally in physical space $M$. In certain circumstances, the dynamics itself can be said to localize physical regions. This is done by identifying regions in physical space with certain submanifolds in configuration space, and determining conditions under which the  projection of  the global dynamics down to these submanifolds coincide the own intrinsic dynamics of the region.

Gauge symmetry of almost any sort complicates our notions of locality. %The reason is that most gauge transformations involve derivatives of the fields, and thus isolating pure gauge components of the fields requires taking inverses of derivatives -- integrals -- which make representations of the physical fields non-local in configuration space.\footnote{If a Fourier transform can be defined, as it happens with a flat metric, then one can indeed represent derivatives,  and possibly their inverses, algebraically in momentum space of course. In more generality, one needs to find eigenfunctions of the derivative operator, analogous to $e^{ik\cdot x}$.  } 
Here I will call a theory kinematically local if the action of the  symmetries  on the configuration variable $\phi(x)$, for a given $x$ depend only the value of $\phi$ on a small neighborhood of $x$. % It is also of fundamental importance, for the constructions of \cite{path_deco} and the gravitational model proposed in \cite{conformal_geodesic}, that the gauge symmetries act as Lie group on configuration space. 
This disallows the standard refoliation symmetry of the ADM form of general relativity, at least for gravitational systems based on spatial metric configuration space.

  The emergent allowed symmetries for metric gravitational theories are discussed in  a third paper,  \cite{conformal_geodesic}. There I construct a  gravitational dynamical system which possesses the right attributes for the formalism to be applicable. It is furthermore a system which possesses the usual transverse-traceless degrees of freedom, but not the usual refoliation invariance of general relativity. 
 
   Once we have a kinematically local dynamical theory,  we will be able to say when the dynamics itself localizes regions, i.e. when the dynamics in one region does not depend on the field content of another region.  This is the notion of dynamical locality that I will pursue in this paper.

In the appendix, I will very quickly introduce basic concepts necessary for our investigations. First, that  of a Jacobi metric. This construction allows us to replace dynamical objects with geometrical ones, in a reasonably straightforward manner. Some of the upshots of having this tool at our disposal are discussed in  \cite{path_deco}.  I will also briefly review the semi-classical approximation I have in mind for these non-relativistic systems.

\section{Configuration space and kinematical locality}\label{sec:observers}
In what follows, I won\rq{}t require specific details of the configuration space.
 I will let $M$ denote a spatial, closed (i.e. compact without boundary)  $n$-dimensional manifold. Typically we will take $n=3$, but  we won't require the dimensionality explicitly. As in \cite{path_deco}, I will take (a given subspace of) the space of sections of some tensor bundle over $M$ to give us the configuration space of the model.  Namely, this could be the space of all vector fields, and/or spinor fields, and/or Riemannian metrics, and/or etc. In general, we can take the fields to be a map which locally on $M$ takes the form $\phi^\alpha:M\rightarrow M\rq{}$, where $M\rq{}$ is another finite-dimensional manifold, and where the subscript denotes values in $M\rq{}$.   I will call the field, or configuration, space over $M$ by the calligraphic $\mathcal{M}$. Each complete determination of the field over $M$ determines a point in $\mathcal{M}$.

 We will also assume that the field equations satisfied by $\phi^\alpha$ are determined by a fixed end-point variation of an action of the form:
 \be\label{equ:general_action}
 S[\phi(t)]=\int dt\, L[\phi^\alpha, \dot\phi^\alpha](t)
 \ee
 where ${L}:T\mathcal{M}\rightarrow C^\infty(M)$,   $\phi^\alpha(0)=\phi^\alpha_o$ and $\phi^\alpha(1)=\phi^\alpha_1$,  and  $\dot\phi^\alpha$ is given by the tangent vector  along the curve (the field history) $\phi^\alpha(t)$, i.e. $\dot\phi^\alpha(t\rq{})=\frac{d}{dt}_{|t\rq{}}\phi^\alpha(t)$.  In most examples, the total Lagrangian can be replaced by the integral of a local Lagrangian density, acting on some $k$-th jet bundle of the field $\phi^\alpha$ and on some $j$-th jet bundle of $\dot \phi^\alpha$.   However, our concept of dynamical locality will not be constrained to theories of such type, in principle applying also to theories that include non-local operators , such as $\nabla^{-2}$.

 In this context, an instantaneous  symmetry acting in configuration space is a transformation $\mathcal{T}_\lambda\cdot \phi^a$ under which $L[\phi^\alpha, \dot\phi^\alpha]$ remains invariant.  That is, we will assume that a Lie group $\mathcal{G}$ (possibly infinite-dimensional) acts linearly on configurations as $\mathcal{T}:\mathcal{G}\times \mathcal{M}\rightarrow\mathcal{M}$. For a general curve $\lambda(t)$ in $\mathcal{G}$,  this action induces an action on $T\mathcal{M}$ in the standard way:
 \be\label{equ:group_action}
 \frac{d}{dt}_{|t=t\rq{}}\left(\mathcal{T}_{\lambda(t)}\cdot \phi^a(t)\right)= \frac{d}{dt}_{|t=t\rq{}}\left(\mathcal{T}_{\lambda(t)}\cdot \phi^a(t\rq{})\right)
+ \frac{d}{dt}_{|t=t\rq{}}\left(\mathcal{T}_{\lambda(t\rq{})}\cdot \phi^a(t)\right) \ee 
 where we are crucially assuming that the group action, $\cdot$\,, is field independent. As we will discuss below, in general relativity, non-spatial diffeomorphisms of space-time do not possess such a representation on (the symplectically reduced) configuration space.\footnote{While it is true that one can extend phase space to include so-called \lq\lq{}embedding variables\rq\rq{}, for which one can better represent the spacetime diffeomorphisms canonically \cite{Isham_parametrized}, but these require further gauge-fixing and the complication of more fields, and it is not clear to me that they fully alleviate the problem.  }
 
These are much more stringent conditions than requiring the action \eqref{equ:general_action} to remain invariant under a field transformation, but they are necessary to form a well-defined quotient.  
  
\subsection{Example: basic structure of Riem.}

For the non-relativistic gravitational systems in consideration here, one could  take configuration space of pure gravity to be the space Riem$(M)=\mathcal{M}$, of positive-definite sections of the symmetric covariant tensor bundle $C_+^\infty(T^*M\otimes_S T^*M)$ over $M$, which forms a subspace (a cone) of the Banach vector space $\mathbb{B}:=C^\infty(T^*M\otimes_S T^*M)$. This linear structure allows us to transport many of the usual finite-dimensional concepts, such as the exponential map and exterior calculus (e.g. Cartan\rq{}s magic formula) to the infinite-dimensional field context.\footnote{See\cite{Michorbook} for conditions under which we can extend these usual theorems to the infinite dimensional Banach context.}

The $\mathcal{M}$ subspace of $\mathbb{B}$ has a one-parameter family of natural Riemannian structures, induced pointwise by the metric $g_{ab}$:
\be\label{equ:supermetric} (v, w)_g:=\int d^3 x\sqrt{g}\, G_\lambda^{abcd}v_{ab}w_{cd}
\ee
where $G_\lambda^{abcd}:=g^{ac}g^{bd}-\lambda g^{ab}g^{cd}$ (when acting on symmetric tensor fields, $v_{ab}=v_{(ab)}$ and so on), and $0\leq\lambda\leq 1/3$. 
These are called the DeWitt supermetrics (with DeWitt value $\lambda$). 

 The supermetric given in \eqref{equ:supermetric} can always be taken as an auxiliary supermetric, but it has little physical content. For instance,  its geodesics have little to do with Einstein's evolution equation (see \cite{Freed} for a characterization of these geodesics). In this paper we largely focus on dynamical systems which we call `of Jacobi type'. Extremal paths for the dynamics of such systems coincide with geodesics with respect to some supermetric in configuration space.  In \cite{conformal_geodesic} I propose such a dynamical system with a given symmetry content, without refoliation invariance but still the usual transverse traceless physical degrees of freedom. The metric of these Jacobi type systems (wrt which the gauge orbits will also be Killing directions) can  be used in place of the auxiliary DeWitt metric to compare configurations. 

As an example of the infinite dimensional gauge symmetries, we could take the spatial diffeomorphisms of $M$,  $\mathcal{G}=$Diff$(M)$. The group action is given by the pull-back, for $(f,g_{ab})\in\mbox{Diff}(M)\times \mathcal{M}$, 
\be\label{diff_transf} (f\,, g_{ab})\mapsto f^*g_{ab}
\ee 
In coordinates, say $f$ goes from a $x^{\alpha\rq{}}$ coordinate system to an $y^{\alpha}$ one,  
$$f^*(g_{\alpha\beta}dy^\alpha dy^\beta (f(x))=(g_{\alpha\beta}\frac{\partial y^{\alpha}}{\partial x^{\alpha\rq{}}}\frac{\partial y^{\beta}}{\partial x^{\beta\rq{}}})dx^{\alpha\rq{}}dx^{\beta\rq{}}(x)$$
with an infinitesimal action given by the Lie derivative, i.e. for a one-parameter group of diffeomorphisms 
\be\label{diff_transf_inf}  \frac{d}{dt}_{|t=t\rq{}}( f(t)^*g_{ab})= f(t\rq{})^*\frac{d}{dt}_{|t=t\rq{}}(f(t)\circ f^{-1}(t\rq{}))^*g_{ab})=f(t\rq{})^*\pounds_\xi g_{ab}
\ee
  for $\xi^a$ the left invariant vector field flow of $f$ at $f(t\rq{})$, i.e. one takes the vector at the identity and transports it by $f^{-1}(t\rq{})$.  If $f(t\rq{})=\mbox{Id}$, then $\frac{d}{dt}_{|t=t\rq{}}\mbox{Fl}(f(t))=\xi^a$, as expected.   Thus: 
 \be\label{diff_transf_inf_metric}  \frac{d}{dt}_{|t=t\rq{}}( f(t)^*g_{ab}(t)) =f(t\rq{})^*(\dot g_{ab}+\pounds_\xi g_{ab})
\ee 
%Suppose that $h(s)\in \mbox{Diff}(M)$, such that $h(0)=f(0)=\mbox{Id}$, then: 

\subsection{Kinematical locality: gravity and diffeomorphisms}

 \paragraph*{Aims of the construction: Product configuration submanifolds}
 One can also define the configuration space $\mathcal{M}$ if $M$ has boundaries, with appropriate (Dirichlet) boundary conditions \cite{Fischer}. We will use the boundary conditions by restricting our attention to a region of $\mathcal{M}$ for which $\phi(x)$ is given over a two dimensional closed surface, which for example could be a two-sphere.\footnote{\label{footnote}I will not discuss much how this surface would be determined under practical circumstances. One could say it is defined locally in $\mathcal{M}$ by a radius of length $r_o$ (according to some local scale) from a given preferred point $x_o$, but then one would have to determine how this point is defined and so on. Suffice it to say here that an observer (whatever that means) can ascertain that there is a region (diffeomorphic) to $S^2$ where the fields stay approximately constant so far as the observer can tell.} 
 
 Let  $S$ be a closed $n-1$ dimensional manifold, such that there is an embedding $\imath:S\hookrightarrow M$, with $\imath(S)$.  The embedding of $S$ defines two regions,  the `interior' $O$ and the exterior $N:=M-(O-\partial O)$, which share the boundary $\partial O=\partial N=S$ (remembering $\partial M=\partial S=0$), where I have identified $\imath(S)$ with $S$ for notational convenience. 
 
   The  manifold $N$, diffeomorhic to $M-O$, will represent in our example the spatial support of the part of the field we don't have physical access to, whereas $O$ is that region of space within $\partial O$ which we do have access to. $O$  is our \lq\lq{}laboratory\rq\rq{}. 
   
   Using the underlying linear space structure of $\mathcal{M}$, our aim will be to  locally form a product structure
    $\mathcal{M}_M^{\partial O}=\mathcal{M}_O^{\partial O}\times \mathcal{M}_N^{\partial O}$, for configuration corresponding to regions and respecting boundary conditions. These boundary conditions will be geometrical  (i.e. not dependent on the representative of the metric). 

   The claim is that for a kinematically local theory,  for a symmetry acting on these configurations as $\mathcal{T}:\mathcal{G}\times \mathcal{M}\rightarrow\mathcal{M}$,  and for any $\lambda\in \mathcal{G}$,  and $\phi\in\mathcal{M}_M^{\partial O}$,  %with the fixed boundary value $\mathcal{T}_\lambda\partial\phi$
   one can  transpose properties of configuration space  to reduced configuration space:
   \be\label{equ:general_split}\frac{\mathcal{M}_M^{\partial O}}{\mathcal{G_M}}\simeq \frac{ \mathcal{M}_O^{\partial O}}{\mathcal{G}_O}\oplus \frac{ \mathcal{M}_N^{\partial O}}{\mathcal{G}_N}\ee with intrinsically defined $\mathcal{G}$.

   \paragraph*{Preamble: Principal fiber bundles} 
   Principal fiber bundles are useful in the discussion of symmetry groups. 
   
  A principal fiber bundle is a manifold $P$, on which a Lie group $G$ acts freely: $P\times G\rightarrow P$,  here we will assume it acts on the left,  $(p, g)\rightarrow g\cdot p=L_g(p)$. The space $\{g\cdot p\,~|~g\in G\}$ is called the the fiber (through $p$).  Identifying $p\sim g\cdot p$, we have a projection operator onto the quotient space $\pi: P\rightarrow P/G$. We will use the square bracket notation to denote the orbit of $p$, i.e.  $[p]=\pi^{-1}(\pi(p))\subset P$, but since there is a one to one correspondence between orbits and points in $P/G$, we sometimes abuse notation and write $[p]=\pi(p)\in P/G$. 
  
  For some open region $U\in P/G$, with the use of a section $\chi:U\rightarrow P$, we can trivialize the bundle: $\pi^{-1}(U)\simeq \chi(U)\times G$, which allows us to write $P\ni p= (x\,,\, g)$ for $x\in U$. Sections intersect each orbit of the gauge group only once. Given two sections, they are always uniquely related by a function $\lambda:U\rightarrow G$, as in $\chi_1(x)= \lambda(x)\cdot\chi_2(x)$.

Given $\mathfrak{v}\in \mathfrak{g}$,  where $\mathfrak{g}:=T_{\mbox{\tiny Id}}G$, we define the fundamental vector field associated to it as  $T_pP\ni\mathfrak{v}_p^\#:=\frac{d}{dt}_{|t=0}\exp{(t\mathfrak{v})}\cdot p$. The vertical subspace  $V_p\subset T_pP$ is the tangent space to the fiber at $p$, i.e.  $V_p=\mbox{span}\{\mathfrak{v}_p^\#\,~|~\mathfrak{v}\in \mathfrak{g}\}$.
  
 In standard gauge field theory, the space of physically equivalent field configurations requires that one quotient the total space of field configurations by the total group of gauge transformations, $\mathcal{A}/$Gau$(G)$, where $\mathcal{A}$ is the field space of connections on space-time,  $G$ is the internal symmetry group and Gau$(G)$ is the group of space-time dependent gauge-transformations. A choice of gauge is a smooth embedding 
 \be\label{equ:gauge_section} \chi: \mathcal{A}/\mbox{Gau}(G)\rightarrow \mathcal{A}
 \ee 
 such that each orbit only intersects the image of $\chi$  once, and it does so transversally, i.e. $\mbox{Im}(T_A\chi)\oplus V_A=T_A\mathcal{A}$. 
 
 Usually, due for instance to Gribov ambiguities, there is no global section $\chi$, and we need to restrict the section to an open set, 
 $$\chi: \pi(\mathcal{U})\rightarrow \pi^{-1}(\pi(\mathcal{U}))$$ where 
 $\pi:\mathcal{A}\rightarrow \mathcal{A}/\mbox{Gau}(G)$ is the projection operator onto the quotient space and $\mathcal{U}$ is some open region in $\mathcal{A}$.  

 \paragraph*{Observer dependent diffeomorphisms} 
 In the case of gravity, we  have according to the above, $P\rightarrow\mathcal{M}=\mbox{Riem}(M)$, with tangent space given by $T_g\mathcal{M}\simeq C^\infty(T^*M\otimes_S T^*M$. The symmetry  group is $G\rightarrow\mbox{Diff}(M)$, with Lie algebra $\mathfrak{g}\rightarrow C^\infty(TM)$ -- the space of vector fields over $M$ -- with Lie algebra given by the vector field commutator. The orbit is given by $\{f^*g_{ab}~| ~ f\in \mbox{Diff}(M)\}$, and finally the vertical space is $V_g=\{\pounds_\xi g_{ab}~|~\xi^a\in C^\infty(TM)\}$, according to \eqref{diff_transf} and \eqref{diff_transf_inf} .

However, in the case of  the field space of gravity mentioned above,  the quotient $\mathcal{M}/$Diff$(M)$ is not even a manifold.   This occurs because certain field configurations have a non-trivial stabilizer subgroup of Diff$(M)$, i.e. non-trivial Iso$_g\subset\, $Diff$(M)$ such that for $f\in\,$Iso$_g$, $f^*g=g$. This disrupts the manifold structure of the quotient space.  

It also means that $\mathcal{M}$ is not a proper principal fiber bundle under the action of the spatial diffeomorphisms.  To remedy this, one can restrict attention to diffeomorphisms that fix a given point $x_o\in M$, and a linear frame (a triad) $\{e_a\}\in C^\infty(L(TM))$, i.e. $f(x_o)=x_o$ and $f^*(e_a)(x_o)=e_a(x_o)$ (see \cite{Giulini}).  I will call these diffeomorphisms $\mbox{Diff}_{x_o}(M)$. We can see $x_o$ should stand physically for the idealized present position of the observer, and in this sense, this subgroup is observer-dependent.  

Ebin and Palais have shown that $\mathcal{M}$ has a local slice, as above \cite{Ebin, Palais}. They did this through the use of the normal exponential map to the orbit along a given point. That is, for an arbitrary given metric $\bar{g}_{ab}$, the orbits were shown to be embedded manifolds, and the normal exponential (according to the supermetric \eqref{equ:supermetric} with $\lambda=0$): $\mbox{Exp}_{\bar{g}}:W\subset V_{\bar g}^\perp\rightarrow \mathcal{M}$ was shown to be a local diffeomorphism onto its image for a given open set $W$,  where, as can easily be seen from the form of $V_g$ and \eqref{equ:supermetric},  
\be\label{vertical_perp}V_{\bar g}^\perp=\{u^{ab}\in T_g\mathcal{M}~|~\bar\nabla_a u^{ab}=0\}
\ee 
 By then showing that the tubular bundle around this orbit was locally diffeomorphic to $\mathcal{M}$, one has, for $g_{ab}\in \pi^{-1}(\pi(\mbox{Im}(\mbox{Exp}_{\bar{g}}(W)))$, a unique $f_g\in \mbox{Diff}(M)$ such that 
$$ f^*_g g_{ab} =\mbox{Exp}_{\bar g}(w_g)
$$ for a unique $w_g\in W$, 
$$w_g=\mbox{Exp}_{\bar{g}}^{-1}(f_g^*g_{ab})$$  Thus 
 $$g_{ab}=f^*_g(\mbox{Exp}_{\bar{g}}(w_g))$$
Furthermore, for $g^2_{ab}=h^*g^1_{ab}$, we then have $f_{g^2}=h^{-1}\circ f_{g^1}$. Thus $w_g=w_{[g]}$ and for any $\tilde{g}_{ab}\in [g_{ab}]\subset\pi^{-1}\pi(\mathcal{U})$, the section is given by 
\be \chi(\pi(\tilde g_{ab}))= \mbox{Exp}_{\bar{g}}(w_{[\tilde{g}]})
\ee In more heuristic terms, $\chi$ takes any  metric along the orbits and translates it along the orbit until it hits the orthogonal  exponential section at the height of $\bar{g}_{ab}$. This intersection gives us the value of $\chi$ for the given equivalence class. 
%$$ f^*_{g^2} g^2_{ab} = f^*_{g^2} h^*g^1_{ab} =\mbox{Exp}(w_g)$$

\paragraph*{Product configuration manifolds.}
Now, let $O$ be given by some embedding $\imath$ of  the (topological) ball $B^3$ into $M$, with  $S^2=\partial B^3$ and  $O=\imath(\mathring{B}^3)\subset M$, such that $x_o\in O$. This is our \lq{}laboratory\rq{} region. 
For a given instantaneous metric $\bar{g}_{ab}$, we define the induced metric on the boundary  $\partial {O}$, \footnote{We should have written $\partial \overline{O}$, where the over bar \emph{is} the closure, but we avoid it in order to not clutter notation and not to confuse it with $\bar{g}_{ab}$, where the bar has nothing to do with closure of course.  } by  $\bar g^{\partial O}_{ab}:={({\bar g}_{ab})}_{|\partial O}$ (or just $\imath^*(\bar{g}_{ab})$).

  I will call the \emph{pre-observer configuration space}, related to the boundary value $\bar g^{\partial O}_{ab}$: 
   \be\label{equ:pre_observer}\mathcal{M}^{\partial O}_M:=\{g_{ab}\in \mathcal{M}\,~ |\,~ (f^*_g g_{ab})_{|\partial O}=\bar g^{\partial O}_{ab}\}
 \ee   
This means that I am fixing \emph{the geometry} on the boundary, not the metric.   For the infinitesimal version, consider the one-parameter family of metrics $g_{ab}(t)\in \mathcal{M}_{\partial}$. Then taking the time derivative, we get from \eqref{diff_transf_inf_metric}, 
$$ f^*_{g(0)}( \dot g_{ab}+\pounds_\xi g_{ab})_{|\partial O}=0
$$
 which means that the metric on the boundary can only vary by an infinitesimal diffeo, $   {(\dot g_{ab})}_{|\partial O}=-{(\pounds_\xi g_{ab})}_{|\partial O}$ , i.e. \emph{the geometry} doesn\rq{}t change.

Now, let $g^O_{ab}\in \mathcal{M}_O:=C^\infty_+(T^*O\otimes_S T^*O)$, with group $\mbox{Diff}_{x_o}(O)$ and algebra defined by $\xi^a\in C^\infty(TO)$, without any boundary conditions   on $\partial O$. It is easy to see that we can trivially embed $\mathcal{M}_O\hookrightarrow \mathcal{M}$.\footnote{Given the characteristic function $\chi_A$ valued in $\{0,1\}$, whose value is:
$ \Theta_A(x)=
\left\lbrace\begin{array}{rl}
1 &\, ~\mbox{if}\, ~x\in A\\
0 &\, ~\mbox{if}\, ~x\notin A
\end{array}\right.
$,  we have that the metric is given  by $ g_{ab}(x):=g_{ab}(x)\Theta_O(x)+g_{ab}(x)\Theta_{O^C}(x)$.  To actually have the embedding explicitly, one could use the regular value theorem with projection from $\mathcal{M}$ to $\mathcal{M}_O$, with such characteristic functions. }  

However, so far, region $O$, and $\mathcal{M}_O$ and $\mathcal{M}_O/\mbox{Diff}_{x_o}(O)$ have very little local meaning, in spite of our restriction to a region $O$.  That is because we have absolutely general metrics in a region diffeomorphic to $B^3$, and thus there is little but topological information. The induced action of a  diffeomorphism of $M$ on $O$ is always a region diffeomorphic to $O$, and since we have all possible metrics on $O$, the action of $\mbox{Diff}_{x_o}(M)$ on $\mathcal{M}$ restricts suitably on $O$ to an action that can be interpreted as one of   $\mbox{Diff}_{x_o}(O)$ on $\mathcal{M}_O$. 

Since the vertical condition \eqref{vertical_perp} is defined locally on $M$, we can go through the same procedure to define a section for $\mathcal{M}_O/\mbox{Diff}_{x_o}(O)$, replacing $M$ by $O$ everywhere; $g_{ab}$ by $g^O_{ab} $ and so on. 
As a first step towards localization, we then can implement the covariant (or geometric) boundary conditions of \eqref{equ:pre_observer} in the space $\mathcal{M}_O$:
 \be\label{equ:pre_observer_O}\mathcal{M}^{\partial O}_O:=\{g_{ab}^O\in \mathcal{M}_O\,~ |\,~ (f^*_{g_O} g^O_{ab})_{|\partial O}=\bar g^{\partial O}_{ab}\}
 \ee   

 To glue together 
 \be\label{glue}\mathcal{M}^{\partial O}_M=\mathcal{M}^{\partial O}_O\oplus \mathcal{M}^{\partial O}_N\ee
  where $N=M-O$, we need to show that  for a metric $g_{ab}\in \mathcal{M}^{\partial O}_M$, for which
 $$g_{ab}(x)=g_{ab}^O(x)\Theta_O(x)+g_{ab}^{N}(x)\Theta_{N}(x)$$
  we have that 
 $$f^*_g g_{ab}(x)=f^*_{g_O}g^O_{ab}(x)\Theta_O(x)+f^*_{g_{N}}g^{N}_{ab}(x)\Theta_{N}(x)
 $$
 The only obstacle one might have for this relation is that the sub-bundle of $T\mathcal{M}$, the space  $V^\perp$ given in \eqref{vertical_perp}, is given implicitly by a differential equation. However, the operator is of first order, only depends on the metric, and is homogeneous (i.e. sourceless). We have that for a tensor $u^{ab}(x)$, $x\in M$, solving $\bar\nabla_a u^{ab}=0$, the solution inside $B^3$ with fixed Dirichlet boundary conditions on the boundary can depend only on the values of the metric inside $B^3$ \cite{Trudinger}. Furthermore, the explicit exponential map of the supermetric, calculated explicitly in \cite{Gil-Medrano} Theorem 3.3, is ultralocal in the metric and initial metric velocity. 

 Thus, since the spaces in equation \eqref{glue} are defined  covariantly with respect to their boundary conditions,    we  finally obtain that  the quotient also splits,
\be\label{equ:quotient_split}
\frac{\mathcal{M}^{\partial O}_M}{\mbox{Diff}(M)}\simeq \frac{\mathcal{M}^{\partial O}_O}{\mbox{Diff}(O)}\oplus \frac{\mathcal{M}^{\partial O}_N}{\mbox{Diff}(N)}
\ee
with every component defined intrinsically.\footnote{  The same would occur if we were to consider Weyl transformations of the metric, since they also act locally and intrinsically on the metric, forming an equivalence relation.  }

%For spatial diffeomorphisms, this means that we c take the support of the smearing of the constraints (i.e. the support of the constraint), $\mbox{supp}(\xi^a)\subset O$ 

\paragraph{A counter-example: refoliations.}
In \cite{Wald_Lee}, using the covariant symplectic formalism, Wald and Lee studied how spacetime difffeomorphisms acting on the field space of general relativity would project down to a phase space related to the choice of a given Cauchy surface. The procedure gave a precise translation between local symmetries acting on field space and constraints acting on phase space.  It was found that the action of non-spatial diffeomorphisms (refoliations) on the entirety of field space  could not be represented as a constraint -- only if one restricted oneself to the subspace of spacetime field configurations which satisfied the equations of motion could one get a representation on phase space. 

The Hamiltonian constraint in the 3+1 ADM form of general relativity is given by 
\be\label{equ:Hamiltonian} \mathcal{H}(x)=\left(R\sqrt{g}-\frac{\pi^{ab}\pi^ {ab}-\frac{1}{2}\pi^2}{\sqrt{g}}\right)(x)=0
\ee where $\pi^{ab}$ is the conjugate momenta to the 3-metric, and one uses the abbreviated notation: $\sqrt{g}=\sqrt{\det{g}}$. By the above remarks, it only generates refoliations of space-time \emph{on-shell}.

Now, in the Hamiltonian framework, the transformations induced by \eqref{equ:Hamiltonian} act infinitesimally on the spatial metric schematically as:
\be\label{equ:refoliations_metric}
\frac{d}{dt}_{|_{t=0}}\mathcal{T}_{\lambda_t} g_{ab}(x)=\lambda\rq{}  (x) \pi_{ab}(x)
\ee
where $\lambda\rq{}$ represents an infinitesimal gauge-parameter (the lapse, in usual nomenclature). Thus possible equivalence relations will depend on the momenta as well  (whether it is the momentum or time derivatives of the metric is not important for this discussion).  To relate $g^1_{ab}$ to  $g^2_{ab}$ we need to be able to integrate a differential equation depending parametrically on $\dot{g}_{ab}$. This is a smooth curve in Riem connecting $g^1_{ab}$ to  $g^2_{ab}$, for a  given initial condition $\dot g_{ab}^1$ and a given \lq\lq{}gauge\rq\rq{}  parameter $\lambda(t)$. 

Now, suppose $g^2_{ab}\sim g^1_{ab}$ and $g^2_{ab}\sim g^3_{ab}$  according to some $\dot g_{ab}^1$,  and $\lambda_1(t)$, and  $\dot g_{ab}^3$ and $\lambda_2(t)$. Let\rq{}s call the curve that solves the equations of motion for the metric (with zero shift) between $g^2_{ab}$ and $g^1_{ab}$ with these conditions, $\gamma_1(t)$, and resp, $\gamma_2(t)$ for the curve obeying the analogous conditions between $g^2_{ab}$ and $g^3_{ab}$.  Since   $g^1_{ab}\sim g^3_{ab}$ only if there exists a solution curve connecting the two, we would only have the transitive property if the opposite of the initial momenta of the solution curve from $g^2_{ab}$ to $g^1_{ab}$, at  $g^2_{ab}$ is the same as the initial momenta for the curve connecting $g^2_{ab}$ and $g^3_{ab}$. I.e. if
\be\label{equ:refoliations_metric}
-\frac{d}{dt}_{|_{t=0}}\mathcal{T}_{\lambda^1_t} g^2_{ab}(x)=-\lambda_1\rq{}  (x) \pi^{1}_{ab}(x)= \frac{d}{dt}_{|_{t=0}}\mathcal{T}_{\lambda^2_t} g^2_{ab}(x)= \lambda_2\rq{}  (x) \pi^{2}_{ab}(x)
\ee
This is what is meant by saying that the Hamiltonian constraint in ADM gravity \cite{ADM} generates the \lq\lq{}groupoid\rq\rq{}  of refoliations, only on-shell. It means we can only have an equivalence relation under very special circumstances -- of equality between fields which do not  depend exclusively on the configurations $g_{ab}^1,g_{ab}^2, g_{ab}^3$ themselves.  Thus we cannot take the quotient as was done in the previous case, and finally, cannot obtain \eqref{equ:general_split} in this case.

In other words, the kinematical gauge structure of the theory -- given before one takes into account the equations of motion, but taking into account the gauge symmetries -- already implies non-locality in configuration space. What shape dynamics \cite{SD_first} does is to find a theory that does have a kinematically local structure, since its symmetries have the properties that allow \eqref{equ:general_split}. However, it is not trivial to see that its dynamics generate local evolution. Thus the purpose of the remaining of the paper is to elaborate conditions under which locality of the dynamics is not postulated, but can emerge. From now on, assuming that the kinematical structure of the theory is local in the sense that it obeys \eqref{equ:general_split}, I will disregard the action of symmetries and concentrate on the dynamics. 

\section{Semi-classical entanglement and dynamical locality.}\label{sec:locality}

There are two aspects of non-locality that we would like to address here. One can be said to be more ``classical": the very equations of motion of the theory incorporate some symmetry that implies observables are non-local. 

The quantum aspect was  already summarized by  Schr\"odinger in \cite{Schroedinger}. In response to the original EPR paper, he pointed out that non-locality is merely a consequence of the non-factorizability (or entanglement) of the two-particle wavefunction: 
\be \label{entanglement}\psi(x_1,x_2)\neq \xi(x_1)\xi(x_2)
\ee
In the present interpretation however, there is a  preferred foliation implicit in the choice of configuration space. A given path in configuration space contains the configuration of both particles in an EPR pair, and thus correlates them.\footnote{Note however, that since in the present \lq\lq{}many-worlds\rq\rq{} type of view, there isn\rq{}t a single history of the Universe, and there can be interference among distinct paths in configuration space. Thus this notion that EPR pairs are correlated in \emph{each} path in configuration space does not contradict Bell\rq{}s theorem.} The only difference between the study of classical non-locality and quantum non-locality, is that the first is restricted to paths that obey the equations of motion.  But it is clear that for the great majority of  paths in configuration space (i.e. those contained in the path integral) do \emph{not} obey the equations of motion and are generically non-local. 

Let us  study  cases where the -- possibly exceptional -- local behavior can be said to emerge, classically, or semi-classically.

\subsection{Dynamical locality }\label{sec:localization}

  From  section \ref{sec:observers}, we  know that there is a natural projection from the space of configurations matching $\partial\phi$ on $\partial O$ to $\mathcal{M}_O^{\partial O}$, i.e. 
  \be\label{equ:proj}\mathsf{pr}_{\mbox{\tiny O}}:\mathcal{M}_M^{\partial O}\rightarrow \mathcal{M}_O^{\partial O}\ee
 This projection can be given by the pull-back of the inclusion $\imath:O\hookrightarrow M$, and it is covariant wrt to the action of the symmetry group, since we  have established geometric \lq{}boundary conditions\rq{}  in \eqref{equ:pre_observer}. 
 
       But I have said nothing about the dynamics in $\mathcal{M}$. The first thing one should demand for the approximate dynamical localization for some region of field space,  is that these boundary conditions are in some sense maintained by the dynamics.  Indeed, it is not because one can find kinematically local  product structures over $\mathcal{M}$, which allows us to consider the split  $\mathcal{M}_M^{\partial O}=\mathcal{M}_O^{\partial O}\times \mathcal{M}_N^{\partial O}$, that its dynamics need to be also of product type.
   
    But the obstructions might be even deeper. Fixed boundary conditions can in general imply that the dynamics factorizes   if one already assumes some degree of locality for the dynamics itself. But this is by no means guaranteed, for the dynamics inside $O$ could be \lq\lq{}entangled\rq\rq{} with dynamics outside of $O$. 

For illustrative purposes, let us take the simplest possible counter-example: the spatial manifold is the one-dimensional circle, $M=S^1$, the observer subset $O$, an interval characterized by $\theta\in [\theta_1,\theta_2]$,  $N$ its complement $S^1-O$. We endow $S^1$ with the free scalar field $\phi:S^1\rightarrow \mathbb{R}$, with
$$S=\int dt \int_{S^1} \, ds\,  \phi(\partial_t^ 2-\partial_\theta^2)\phi$$
and so extremal field histories satisfy $\square\phi=0$ (where $\partial^2$ induced from the standard Euclidean metric). The eigenfunctions are $e^{i\mathbf{k}\cdot \mathbf{x}}$ (in relativistic notation), and the allowed eigenmodes are determined by the  boundary conditions. The problem here is that the closed manifold already has periodic boundary conditions, and thus further imposing the local ones,  $\phi(\theta_1,t)=\phi(\theta_2,t)=0$, can be problematic. Since global modes are parametrized by $\frac{n\pi}{[S^1]}$, and the local ones by   $\frac{n\rq{}\pi}{[O]}$, where $[A]$ is the length of the set, if the ratio of lengths of $O$ and $S^1$ is irrational, there is no global solution that reduces to a local solution, as $n\rq{}\neq n\frac{[O]}{[S^1]}, \, \forall n,n\rq{}\in \mathbb{N}$,  and the only solution satisfying all the conditions would be a constant field. We say this would be \lq\lq{} noticeable\rq\rq{}  in $O$, since their inhabitants are not getting all the extremal field histories they might expect intrinsically. Thus the dynamics of $S^1$ in this case does not decouple into a product of the dynamics of each segment.

Of course, the example is artificially made to be stationary, and that is what makes it look \lq\lq{} non-local\rq\rq{}. Any change in boundary conditions would be causally propagated. One could instead write down a Lagrangian for which we get elliptic equations of motion, static by fiat, $\partial^ 2\phi=0$. The only solution for this equation on $S^1$ is constant, but on the annulus  $S^1\times [a,b]$ with appropriate Dirichlet boundary conditions we again obtain similar properties (we illustrate such a solution in the appendix). 

 In any case,  I would like to make statements about locality that can be as general as possible, without regards to the specific form of the Lagrangian (for example, it could be  given implicitly through the inverse of a non-linear polynomial differential equation, as is the case of shape dynamics \cite{SD_first}) and relies instead only on properties of the solutions of the equations of motion, i.e. only on geometric properties of the extremals of the action.  

For an action which is  not necessarily Lorentzian, and not necessarily spatially local,  rewriting \eqref{equ:general_action}, 
$$
 S[\phi(t)]=\int dt\, L[\phi, \dot\phi](t)
 $$
 Let $\phi_i$ and $\phi_f \in \mathcal{M}$ be given initial and final field configurations. An extremal of the action between these fixed points is a field history $\phi^{\mbox{\tiny cl}}(t)$ such that for any two-parameter family of fields $\phi(t,s)\in \mathcal{M}$, such that $\phi(t,0)=\phi^{\mbox{\tiny cl}}(t)$ and $\phi(0,s)=\phi_i$ and  $\phi(1,s)=\phi_f$, we have
 $$\frac{\partial S[\phi(t,s)]}{\partial s}{\Big|_{s=0}}=0
 $$
 This notion can be applied for total Lagrangians which are defined intrinsically for a given region $O$ (which is not necessarily directly related to the restriction of the total Lagrangian on $M$). For instance, in an extreme case, suppose that 
 $$L[\phi, \dot\phi]=\int_M d^3x(\dot \phi^2+\phi\frac{1}{\nabla^2}\phi+j\phi)
 $$
 where $j$ is some source for $\phi$. Now, notice  the appearance of $\nabla^{-2}$. Upon calculating the equations of motion, this will act non-locally on the sources, $j$. Thus even if  $\mathsf{supp} (j)\subset O^C$, the projection of the dynamics of the fields to $O$ clearly does not result in the same dynamics, for any time interval, as if we had just defined the Lagrangian intrinsically, as 
  $$L[\phi_O, \dot\phi_O]=\int_O d^3x(\dot \phi_O^2+\phi_O\frac{1}{\nabla^2}\phi_O+j\phi_O)+\int_{\partial\bar{O}} d^2x\, B(x)
 $$ (where we need to add a boundary term for the variational principle to make sense \cite{RT}).\footnote{In this case, upon an infinite expansion of the inverse Laplacian operator, we would have to add an infinite expansion of derivatives of $\phi$ on the boundary, which is in accord with the theory being non-local.} 
%In this spirit of generality, let me merely assume that a Lagrangian can be defined over any manifold with boundary: $\mathcal{L}:A\rightarrow \mathbb{R}$.  I will take it to be expressible as a (possibly multiple)  integral of a (resp. possibly non-local) volume form (here in the case of two integrals, using the DeWitt mixed functional depencence notation): \be\label{equ:non-local_lagrangian}\mathcal{L}_A[\phi,\dot\phi]=\int_A d^3x\,\int_A d^3y\,\sqrt{g}(x) \sqrt{g}(y)\, \omega[\phi,\dot\phi;x,y)\mu[\phi,\dot\phi;x)\eefor arbitrary (two-point) functional $\omega[\phi,\dot\phi;x,y)$ and $\mu[\phi,\dot\phi;x)$.  

%A particular example with these properties and other desirable ones will be studied in section \ref{sec:toy_model}. Using \eqref{equ:non-local_lagrangian},  we could  have a dynamical system $\mathcal{L}_O$,  intrinsic to $O$, or one intrinsic to $M-O$, and so on. Due to the possible occurrence of multiple integrals, the system might be  non-local. It is not necessarily the case that by restricting the outermost integral $\int_M d^3x\rightarrow \int_A d^3x$, the whole integrand also localizes,  $\mathcal{L}_M\rightarrow \mathcal{L}_A$ . Due to the presence of diffeomorphisms, a characterization of locality becomes  muddled. 

What I will attempt to describe here is how the dynamics over the whole of $M$ relates to an intrinsically defined (e.g. over $O$) dynamical system, which allows us to determine empirically (or dynamically) when a region localizes. That is, I would like to focus on properties of the extremals of the action. 
The way I will define $O$ to be semi-classically localized, will apply \emph{only for some given region (or process) in configuration space}. In other words, the given region $O$ is not absolutely local, it is only local for some given field content. 
 
  Given extremal paths $\phi^{\mbox{\tiny cl}}(t)$ over $\mathcal{M}$, I would like to compare their \emph{restriction}, $\mathsf{pr}_O\phi^{\mbox{\tiny cl}}(t)$,  with \emph{intrinsic} extremal paths $\phi^{\mbox{\tiny cl}}(t)_{O}$ (with the same initial and final projected conditions). Roughly, I will say that the dynamics of $O$ is local for that initial and final field content if  according to \eqref{equ:general_split}, $\frac{\mathcal{M}_{\partial\phi}}{\mathcal{G_M}}\simeq \frac{ \mathcal{M}_O^{\partial O}}{\mathcal{G}_O}\oplus \frac{ \mathcal{M}_N^{\partial O}}{\mathcal{G}_N}$, the image of the projection  of an extremal curve on the global space to the local space is  also a (local) extremal curve, and if it is  a surjective map on the space of extremal curves on $O$. Borrowing notation from quantum mechanics, we  want evolution and projection to the given region to commute:
  \be\label{equ:local_H} e^{i\hat{H}_Ot}\circ \mathsf{pr}_{\mbox{\tiny O}}= \mathsf{pr}_{\mbox{\tiny O}}\circ e^{i\hat{H}_Mt}\ee
 where $e^{i\hat{H}t}$ is the unitary time evolution operator,  the lhs denote evolution for the intrinsic action in $O$, and the rhs denotes projection of the total evolution onto the $\mathcal{M}_O^{\partial O}$ subspace.  

  \paragraph*{Localization and ``domain of dependence".}
   I want to maintain some order of approximation, for boundary conditions are very seldom \emph{exactly} kept intact by the dynamics.  Furthermore, at this level, I will use an abstract definition for a field $\phi$, and without local symmetry groups.  Having \eqref{equ:local_H} in mind,  I thus define:  
\begin{defi}[Semi-classical localization and independent regions.]\label{def:spatial_localization}
A spatial region $O\subset M$, bounded by a codimension 1 compact surface $\partial \bar{O}$ will be said  {\bf semi-classically $\epsilon$-localized} (or independent of $O^C$), between two given partial field configurations $\phi^i_{O}$ and $\phi^f_{O}\in \mathcal{M}^{\partial O}_O$ (with {\bf fixed} boundary conditions $\partial \phi_O$ on $\partial O$), if every extremal curve $\phi^{\mbox{\tiny cl}}(t)\subset\mathcal{M}$ between $\phi^i$ and $\phi^f$ which initially and finally satisfy $\mathsf{pr}_O(\phi^i)=\phi^i_{O}$ and $\mathsf{pr}_O(\phi^f)=\phi^f_{O}$,\footnote{But are not necessarily in $\mathcal{M}^{\partial O}_M$, i.e. not necessarily maintaining the boundary conditions throughout.} obeys the following two conditions: % {\bf i}) $\phi^{\mbox{\tiny cl}}(t)$ is approximately equal to a curve $\bar\phi^{\mbox{\tiny cl}}(t)$ in $\mathcal{M}^{\partial O}_M$, i.e. it approximately maintains the given boundary conditions, 
%\be\label{equ:spatial_local1} \|\phi^{\mbox{\tiny cl}}(t)-\bar\phi^{\mbox{\tiny cl}}(t) \|_{\phi^{\mbox{\tiny cl}}(t)}\leq \epsilon\,  , \forall t\ee
%in the local inner product of $\mathcal{M}$, at $\phi^{\mbox{\tiny cl}}(t)$.
 {\bf i}) $\phi^{\mbox{\tiny cl}}(t)$ restricts to a curve $\mathsf{pr}_O(\phi^{\mbox{\tiny cl}}(t))$ which is approximately extremal on the intrinsic reduced configuration space $\mathcal{M}_O^{\partial O}$ over $O$, i.e.
 \be\label{equ:injective}
 \forall\, \phi^{\mbox{\tiny cl}}(t)\, ,\exists\,  \phi_O^{\mbox{\tiny cl}}(t) \,~|\,~ \mathsf{pr}_O(\phi^{\mbox{\tiny cl}}(t))\approx \phi_O^{\mbox{\tiny cl}}(t)
 \ee and   {\bf ii}) \emph{any} local extremal curve $\phi^{\mbox{\tiny cl}}_O(t)$ is approximated in this manner, i.e. is approximated by the projection of a global extremal curve,  
 \be\label{equ:surjective}
 \forall\, \phi_O^{\mbox{\tiny cl}}(t)\, ,\exists\,  \phi^{\mbox{\tiny cl}}(t) \,~|\,~ \mathsf{pr}_O(\phi^{\mbox{\tiny cl}}(t))\approx \phi_O^{\mbox{\tiny cl}}(t)
 \ee
 %Given a disjoint region $O'\subset M$, $O\cap O'=\emptyset$,  replacing $M$ by $O\dot\cup O'$, we will say instead that $O$ is {\bf independent} of $O'$ between the two given partial field configurations. 
\end{defi}
 
%Condition i) is necessary  because we want to use the inner product of $\mathcal{M}_O^{\partial O}$ in order to establish approximations which would be locally perceivable, and thus we need an object which projects into $\mathcal{M}_O^{\partial O}$, which we can't with  $\phi^{\mbox{\tiny cl}}(t)$ but can with  $\bar\phi^{\mbox{\tiny cl}}(t)$.
 In a general situation,  there could be extremal curves in the entire space that don\rq{}t restrict to intrinsically local extremal curves. The more obvious case in which this could happen is if none of the extremal curves connecting the initial and final point maintain the boundary conditions. Even if they do maintain the boundary conditions, failure of condition ({\bf{i}}) would translate into observers seeing dynamics (projections of the global, true dynamics) that could not be explained by the local laws of the laboratory.   Condition i), or equation \eqref{equ:injective}, tells us that there always exists an extremal curve  $\phi^{\mbox{\tiny cl}}_O(t)$ in $\mathcal{M}_O^{\partial O}$ such that
\be\label{equ:spatial_local2}
 \|\phi^{\mbox{\tiny cl}}_O(t)-\mathsf{pr}_O(\phi^{\mbox{\tiny cl}}(t)) \|_{\phi^{\mbox{\tiny cl}}_O(t)}\leq \epsilon\,  , \forall t
 \ee
 in the inner product of $\mathcal{M}_O$ (since $\mathsf{pr}_O(\phi^{\mbox{\tiny cl}}(t))$ doesn\rq{}t necessarily belong to $\mathcal{M}_O^{\partial O}$)  at $\phi^{\mbox{\tiny cl}}_O(t)$. In the case of gravitational fields,  equation \eqref{equ:spatial_local2} would amount to 
 $$ \| h^{\mbox \tiny {cl}}_{ab}(t)-g^{O\mbox \tiny {cl}}_{ab}(t)\|=\int_O d^3 x\, \sqrt{g^O} \left(h^{ab}h_{ab}-2 h+3\right)(t)\leq\epsilon,\, ~\forall t
 $$ where the integrand is  a positive function for all $x\in O$, and I have defined $h^{\mbox \tiny {cl}}_{ab}=\imath^*g^{\mbox \tiny {cl}}_{ab}$, for $\imath:O\hookrightarrow M$,  $g^{\mbox \tiny {cl}}_{ab}\in \mathcal{M}$, $g^{O\mbox \tiny {cl}}_{ab}\in \mathcal{M}^{\partial O}_O$ and $h=h^{ab}g_{ab}$.
 
 In sum, the condition  tells us that by projecting some global extremal curve onto the local patch, we would  get something that  local observers would be able to explain using their own local data and laws, with no large extraneous influences. 
 
In the general case one might not be able to ignore external influences in the dynamics of $O$, and thus extremal curves (dynamics) calculated in the laboratory without taking into account the remaining of the Universe might turn out to be wrong, because the global dynamics does not project down to the local one. Furthermore,  this could happen even if the boundary conditions are maintained throughout evolution.    Condition ii) is the condition that \emph{all} local extremal curves would be obtainable from the projection, and thus local observers would not see  non-local effects creeping in through some sort of ``censorship" onto their intrinsic equations of motion, for example, if some, but not all of the possible local intrinsic dynamics can be observed (i.e. not all intrinsic dynamics is correlated with global dynamics).

\begin{figure}[h]\label{figure:multiple}
\begin{center}
\includegraphics[width=10cm, height=4.5cm]{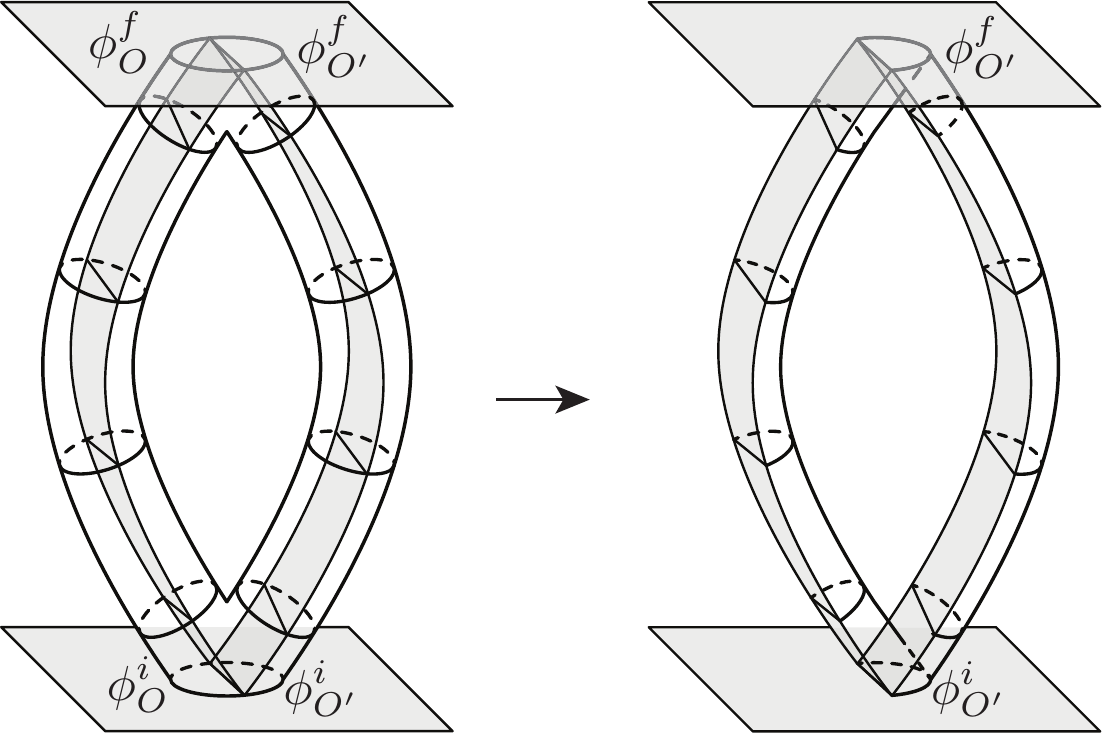}
\caption{ $O\rq{}$ independent of $O$ between $\phi^i$ and $\phi^f$. On the lhs, the tubes represent the extremal field histories of the joint region $O\cup O\rq{}$. On the rhs we see the extremal field histories intrinsic to $O\rq$. The image of the projection of the total histories match the intrinsic histories.  }
\end{center}
\end{figure} 

Definition \ref{def:spatial_localization} implies that if somehow an observer could determine the region where it is at as the one bounded by $\partial O$, semi-classically, this observer would feel no influence from fields existing outside of $O$. She would still be able to observe \emph{any} solution of the equations of motion with the given initial and final conditions as explained \emph{solely} by the fields in $O$, without reference to $M_O$.  It is the observer's ``future domain of dependence", here transported to a possibly non-local, non-relativistic setting.

\paragraph*{Comparison with the relativistic setting.}
\begin{figure}[h]\label{fig:domain}
\begin{center}
\includegraphics[width=10cm, height=4.5cm]{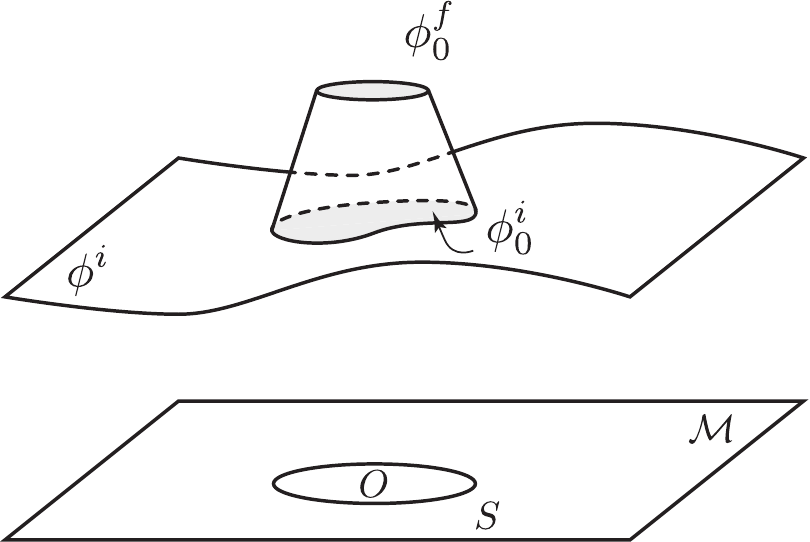}
\caption{The distinction between manifold and submanifold structure (bottom), and initial and final field configurations (top). In the relativistic fields context, the cone between $\phi^i_O$ and $\phi^f_O$ represents (a slice of) the `future domain of dependence' of (the field $\phi^i_O$ at) $O$. }
\end{center}
\end{figure}  
In the usual relativistic context (see figure 2 above), if $O$ is a subset of a smooth Lorentzian manifold, the domain of dependence of $O$ is the set of points $p$ for which all causal curves going through $p$ intersect $O$. To translate this to our context, we need to make some distinctions. First, we need to distinguish between the region $O$ -- just a manifold -- and the initial and final fields at $O$.   In this relativistic context -- which implies  differential equations of motion for perturbations of the metric which are hyperbolic in nature --   an enveloping cone exists in space-time such that metric perturbations  from an initial Cauchy surface don\rq{}t cross it. 

	This is the perturbative statement, that to reach a final partial field configuration  $\phi^f_O$ from $\phi^i_O$, the field configuration outside of $O$ is completely irrelevant. The evolution of the embedded field is thus identical to what it would have been intrinsically, i.e. if $O$ constituted the whole Universe and one had no knowledge of anything beyond $O$.  However, the crucial distinction is that this happens always, independently of field content. I.e. independently of which particular initial and final configuration, independently of the particular dynamics we are considering, a region $O$ will have its domain of dependence.\footnote{If one wanted to translate  definition \ref{def:spatial_localization} to the space-time picture, one would require that the fields not change along the boundary of the region, i.e. that we have a fixed field configuration $\partial\phi$. For metric fields, this would require  Killing vectors $K^\mu$ tangent to the boundary. {A vector field which is  surface-forming, $\nabla_{[\nu}K_{\mu]}=0$, and shear-free, $\nabla_{(\nu}K_{\mu)}$, is also covariantly conserved, $\nabla_\nu K_{\mu}=0$. }} In the non-hyperbolic, non-local case, this is not necessarily true, even for short times and small regions. Nonetheless, we can stipulate conditions under which this would \emph{effectively happen}.   

%\footnote{Also note that the (gauge-fixed) initial and final boundary value problem is different than the Cauchy problem: it would have uniqueness properties for a given small time of evolution, but we could have existence but not uniqueness for longer evolutions, and thus multiple extremal histories between $\phi^i_O$ and $\phi^f_O$. This is related to the `thick-sandwich problem' in general relativity. }  

\subsection{Semi-classical entanglement between regions and ``cluster decomposition".}

Definition \ref{def:spatial_localization} still allows for the  sort of non-local correlations corresponding to semi-classical entanglement between two regions. That is because although extremal paths in $O$ are in some sense "independent" of what occurs in $M-O$, depending only on what happens in $O$ (for the particular field content in question),  extremal paths in $M-O$ might still be correlated with extremal paths in $O$.  It is possible that the global dynamics projects surjectively onto the intrinsic dynamics of $O$, but not onto the intrinsic dynamics of $M-O$. In other words, for two regions $O$ and $O\rq{}$, even if $O$ seems independent from $O'$, $O'$ might be dependent on $O$.

 In the simplest example on the circle $M=S^1$ given above, at the beginning of section \ref{sec:localization}, suppose that the lengths of the intervals are proportional, e.g.  $[O]/[N]=\frac{1}{3}$. All intrinsic eigenmodes  of $O$ (with the given boundary conditions), with $p^O_\phi=\frac{n\pi}{[O]}$ are allowed and obtainable by restrictions of the mode eigenfunctions on $M$, for $p^M_\phi=\frac{n\rq{}\pi}{[M]}$, and $n\rq{}=4n$ (so that for instance the first mode of $O$ is the fourth of $M$)  but we can only re-obtain modes in $N$ which are multiple of the modes in $O$ for our given extra boundary conditions, and thus not all intrinsic solutions of $N$ are obtainable by restrictions of global.  For instance, we would not be able to obtain the fundamental mode of the intrinsic $N$ system, for it lies in between the first and second eigenmode of $M$.

For a more abstract dynamical example, suppose that for all extremal curves $\gamma^\alpha$ between $\phi^i_{O\dot\cup O'}$ and  $\phi^f_{O\dot\cup O'}$, there is a unique extremal curve restricting to $O'$ for each extremal curve restricted to $O$ (e.g. a Bell pair). Using the semi-classical expansion \eqref{equ:semi_classical_exp}, we have, for a gauge-fixed, or reduced action:
\begin{align}\label{equ:maximally_entangled}
 K^{|O\dot \cup O'}_{\mbox{\tiny{cl}}}(\phi^i_{|O}\dot\cup \phi^i_{|O'}, \phi^f_{|O}\dot\cup \phi^f_{|O'})&=A\sum_{\alpha\in I} ((\Delta_\alpha)^{1/2} e^{i S[\gamma^\alpha_{|O}+\gamma^\alpha_{|O'}]})\\
 ~&\neq A\sum_{\beta\in I_O} ((\Delta^{O}_\beta)^{1/2} e^{i S[\gamma^\beta_{|O}]})\sum_{\beta'\in I_{O'}}((\Delta^{O'}_{\beta'})^{1/2} e^{i S[\gamma^{\beta'}_{|O'}]})\nonumber\\
 &= K^{O}_{\mbox{\tiny{cl}}}(\phi^r_{|O}, \phi^f_{O})K^{|O'}_{\mbox{\tiny{cl}}}(\phi^r_{|O'}, \phi^f_{|O'})\nonumber
\end{align}
where $I, I_O, I_{O'}$ parametrize the intrinsically extremal paths in (resp) the joint region, region $O$ and $O'$ and we have denoted $\mathsf{pr}_A(\phi)=:\phi_{|A}$ to simplify notation here.  

In this example,  the partial classical field histories -- which are restrictions of a joint classical field history --  are perfectly correlated to form each joint classical field history.  Heuristically, the distributional density of states on $\mathcal{M}_{|O\dot\cup O'}$ is (semi-classically) peaked on some `diagonal' submanifold of $\mathcal{M}_O^{\partial O}\times \mathcal{M}^{\partial O\rq{}}_{O'}$ and does not factorize.

Now I will show that if the  regions $O_k$ are all are \emph{mutually independent} as per definition \ref{def:spatial_localization}, then we obtain the aimed for semi-classical cluster decomposition. That is:
\begin{prop}\label{prop}
Given regions $O_k\subset M$, $k\in \Lambda$, mutually independent in the configuration space region $\mathcal{U}\subset \mathcal{M}$ according to definition \ref{def:spatial_localization}, from equation \eqref{equ:semi_classical_exp} we have that for $\phi^i, \phi^f \in \mathcal{U}$, 
\be\label{equ:cluster}K(\phi^i_{|\dot\bigcup_k O_k}, \phi^f_{|\dot\bigcup_k O_k})\approx \prod_k K^{O_k}_{\mbox{\tiny{cl}}}(\phi^i_{O_k}, \phi^f_{O_k})\ee 
where $K^{O_k}$ is defined with the intrinsic (gauge-fixed) action on $O_k$, and $\phi^i_{O_k}:=\phi^i_{|O_k}$, the same holding for $\phi^f$. \end{prop}   

{\bf Proof:} If the spatial regions are mutually independent, this means that in a given region of configuration space the on-shell energy functional  has independent  dependences on the different partial field configurations (see equations \eqref{equ:warp_metric} and \eqref{equ:prod_metric} below for the examples in which dynamics is given by a Jacobi metric). For the gauge-fixed (not necessarily on-shell) action functional yielding these extremal paths in the given region $\mathcal{U}$, it also means that 
 \be\label{equ:additivity} S_{|O\cup O'}=A_OS_{|O}+A_{O'}S_{|O'}\, ,\, ~\mbox{for paths in }\, \mathcal{U}\ee where the $A$'s are field independent constants which we will reabsorb into the intrinsic action.\footnote{It could still be true however that these constants don't match for different regions $\mathcal{U}$ where semi-classical decoupling is valid. This is not a big problem however, because the product formula is in any case only shown to hold separately for each of these particular regions. } Thus the choice of initial on-shell momentum in region $O_k$ (analogous to the initial field velocity) will not affect the final configuration in region $O_j$ \emph{for the extremal paths}, yielding: 
$$ \frac{\delta^2 S_{\mbox{\tiny{cl}}}}{\delta\phi^r_{|O_k}\delta\phi^f_{|O_j}}=\left(\frac{\delta\phi^f_{|O_k}}{\delta \pi^r_{|O_j}}\right)^{-1}\approx 0\, ~\mbox{for}\, ~k\neq j$$
%This can be seen by supposing that $\frac{\delta S}{\delta\phi^i_{|O_k}}$ had local support on $\nu^f_j$. That is,  $$S[(\phi^i_{|O_k}+\delta\phi^i_{|O_k})\dot\cup \phi^i_{|O_j}, \phi^f_{|O_k}\dot\cup \phi^f_{|O_j}]- S[\phi^i_{|O_k}\dot\cup \phi^i_{|O_j}, \phi^f_{|O_k}\dot\cup \phi^f_{|O_j}]$$
  and thus also the Van-Vleck determinant factorizes,  $\Delta\approx \prod_{j}\Delta_{|O_j}$.

Now,  rewriting the semi-classical path integral we obtain from \eqref{equ:semi_classical_exp} (compare with \eqref{equ:maximally_entangled}): 
\begin{align}
\sum_{\alpha\in I}(\Delta_\alpha)^{1/2}e^{i S[\gamma^\alpha]}&=\sum_{\alpha\in I}(\Delta^O_\alpha)^{1/2}(\Delta^{O'}_\alpha)^{1/2}e^{i S[\gamma_O^\alpha]}e^{i S[\gamma_{O'}^\alpha]}\nonumber\\
&=\sum_{\beta\in I_O} ((\Delta^{O}_\beta)^{1/2} e^{i S[\gamma^\beta_{|O}]})\sum_{\beta'\in I_{O'}}((\Delta^{O'}_{\beta'})^{1/2} e^{i S[\gamma^{\beta'}_{|O'}]})\label{equ:extremal_decoupling}
\end{align}  
 where, again, $I\,,\, I_O\,,\, I_{O'}$ parametrize the complete set of extremal paths in each region, and   on the passage from the first to the second line we used mutual independence, which implies that $I$ can be reparametrized into  $I_O\,,\, I_{O'}$. This can be seen by splitting $\alpha\in I$, e.g. let $\alpha_{1\beta}$ parametrize all the extremal paths in the total manifold which coincide with extremal path 1 in region $O$, and so on, which implies that using the assumption of mutual independence --   a bijection between extremal paths in the joint region and (the union of) those intrinsic to each region, i.e. between $I$ and $I_O\times I_{O'}$ -- we can write $\sum_{\alpha\in I}=\sum_{\alpha_{\beta\beta'}}$ for   $\beta\, ,\, \beta' \in I_O\,,\, I_{O'}$. Equation \eqref{equ:extremal_decoupling} follows then from the product formula for the Van-Vleck and the additivity of the action in this region of configuration space, \eqref{equ:additivity}.

Finally, generalizing to more than two decoupled regions,  we obtain \eqref{equ:cluster}:   
$$K(\phi^i_{|\dot\bigcup_k O_k}, \phi^f_{|\dot\bigcup_k O_k})\approx \prod_k K^{O^k}_{\mbox{\tiny{cl}}}(\phi^i_{O_k}, \phi^f_{O_k})$$ which means that the amplitude for the arc-length parametrized semi-classical path integral decouples. 
It is our version of "clustering decomposition".  $\square$.

\subsection{Jacobi metric interpretation}

For a Jacobi-type action (see section \ref{app:Jacobi_metric}), definition \ref{def:spatial_localization} implies  the existence of a totally geodesic foliation in Riemannian geometry. Condition ii) is analogous to saying that geodesics of the ambient manifold $\mathcal{M}$ project onto geodesics of the submanifold $\mathcal{M}_O^{\partial O}$, and condition ii) says that all geodesics of the submanifold are obtainable from this projection. {In the usual Riemannian geometry context, the conditions are somewhat different, namely, that any geodesic in each leaf is a geodesic of the ambient manifold as well. In that case, the geometry of the submanifold is given by the induced metric on it, dispensing with the need to project onto the submanifolds. Here we required an extra item in the definition because the submanifolds come endowed with their own (super)metric and dynamics.}\footnote{It is also important to note that definition \ref{def:spatial_localization} goes beyond the demand, in Hamiltonian mechanics, that the dynamics preserve the submanifolds defined by the region. That is because we want to say that the region has an intrinsically defined dynamics, which must match, and be given by, the projection of the Hamiltonian flow. More effort needs to be put into the translation of these concepts to the Hamiltonian (symplectic geometry) setting. }  

 As we saw above, we have  natural local product structure  $\mathcal{M}_M^{\partial O}\simeq \mathcal{M}_O^{\partial O}
 \times \mathcal{M}_N^{\partial O}$.
Now, take a region $\mathcal{V}$ of $\mathcal{M}$ close to $\mathcal{M}_M^{\partial O}$,  in the norm given by the configuration space supermetric \eqref{equ:supermetric}. Suppose that there is a neighborhood $\mathcal{U}_O$ of $\mathsf{pr}_O\phi^i=:\phi^i_{O}$ and $\mathsf{pr}_O\phi^f=:\phi^f_{O}$  for which definition \eqref{def:spatial_localization} holds (and thus necessarily $\phi^i\,,\,  \phi^f\, \in \mathcal{V}$).  Definition \ref{def:spatial_localization} means that ambient extremal paths, i.e. paths in $\mathcal{V}\subset \mathcal{M}$,  between the two given points $\phi^i_{O}=\mathsf{pr}_O\phi^i$ and $\phi^f_{O}=\mathsf{pr}_O\phi^f$  are (approximately) also intrinsically extremal, i.e. extremal in $\mathcal{M}_O^{\partial O}$. 

Let us now sketch the analogy between our notion of locality and that of a totally geodesic foliation. For a subset $\mathcal{U}\subset \mathcal{V}$ with leaves in $\mathcal{U}_O$ and the assumed Riemannian metric structure of configuration spaces given by the Jacobi metric. 

   Thus if a neighborhood $\mathcal{U}_O\subset \mathcal{M}_O^{\partial O}$ as described above exists, we have a totally geodesic \emph{foliation} of \be\label{equ:local_prod_set}\mathcal{U}:=\mathsf{pr}^{-1}_{\mbox{\tiny O}}(\mathcal{U}_O)\cap \mathcal{V}\ee
  
  In turn, it is not difficult to show, but lies beyond the scope of this paper, that, \emph{in the finite dimensional case, with compact leaves}, a totally geodesic foliation has locally isometric leaves \cite{geodesic_foliation}, i.e. the metric on the leaves is independent of the point along the fiber (the direction transversal to the fiber). Let us assume that this holds also for our infinite-dimensional context. The product structure then allows us to decompose the metric on $\mathcal{U}$ given in \eqref{equ:local_prod_set} as a warped product: 
   \be\label{equ:warp_metric}\langle\, \cdot\, ,\, \cdot\rangle= \langle\, \cdot\, ,\, \cdot\rangle_{\mathsf{pr}_{\mbox{\tiny O}}(\mathcal{U})}+F_{\mathsf{pr}_{\mbox{\tiny O}}(\mathcal{U})}\langle\, \cdot\, ,\, \cdot\rangle_{\mathsf{pr}_{\mbox{\tiny N}}(\mathcal{U})}\ee
   where $F_{\mathsf{pr}_{\mbox{\tiny O}}(\mathcal{U})}:\mathsf{pr}_{\mbox{\tiny O}}(\mathcal{U})\rightarrow \mathbb{R}$, and the two products act on tangent spaces  of $\mathcal{M}_O^{\partial O}$ and $\mathcal{M}_N^{\partial O}$ respectively. In other words, one can look at the dynamics of (a proper subset of) $\mathcal{U}_O$ according to its own metric, without regards to what is happening non-locally.
   
   Before we end this section, let me call attention to the fact that the warped product equation \eqref{equ:warp_metric} makes it clear that  extremal paths for $N$ could still very well depend on which extremal paths are followed in $O$, that is, the dynamics in the two regions could still be (non-locally) correlated, but this correlation would only be felt in the dynamics of $N$.   

If two regions are independent, we obtain from \eqref{equ:warp_metric}, the product metric: 
\be\label{equ:prod_metric}
\langle\, \cdot\, ,\, \cdot\rangle= \langle\, \cdot\, ,\, \cdot\rangle_{\mathcal{U}_O}+\langle\, \cdot\, ,\, \cdot\rangle_{\mathcal{U}_{O'}}\ee
and thus extremal paths of the joint region are also of product type. This case would streamline the proof of proposition \ref{prop}.
%This heuristically implies that in the semi-classical approximation, \be\langle \phi^i|e^{i\hat{H}t}\circ \mathsf{pr}_{\mbox{\tiny O}}|\phi^f\rangle=\langle \phi^i| \mathsf{pr}_{\mbox{\tiny O}}\circ e^{i\hat{H}t}|\phi^f\rangle\ee
\subsubsection*{Effective separation of modes}

Let us sketch an application of  the independence criteria above to modes, or scales, as opposed to spatial regions. Now, $\mathcal{M}$ is a subspace of a vector space (which is not itself a vector space). Let us suppose for definiteness that $M=S^3$. In principle we could describe any point of $\mathcal{M}$ in terms of  spherical harmonics. However, to have a separation that has any physical meaning, it is better to use a mode decomposition that is specific to the given neighborhood of $\mathcal{M}$ we want to apply it to. 

For instance, in a neighborhood of a base point (or background metric) $g_o$, we can use the eigenmodes of the (tensor) Laplacian $\nabla$. That is, let $h_i\in C^\infty(TM\otimes_S TM)$ be such that:  
\be\label{equ:eigenbasis}
\nabla h_i=\lambda_i\, h_i
\ee
 Tensors in the vector spaces  $\mathtt{span}\{h_i\}_{i\in \Lambda_1}$ and $\mathtt{span}\{h_i\}_{i\in \Lambda_2}$ might also (approximately) decouple (e.g. around some Jacobi radius from $g_o$). In principle this decoupling would have a similar interpretation as in definition \ref{def:spatial_localization}, but we have not worked out in detail how the picture in this case is supposed to work. It might be then possible to write the amplitude kernel between an initial configuration $\phi_i=\lambda_i^jh_j$ and a final one $\phi^f=\sum\lambda_f^jh_j$ as: 
\be  K(\sum\lambda_i^jh_j,\sum\lambda_f^jh_j )\approx K_{\mbox{\tiny{cl}}}(\sum_{j\in \Lambda_1}\lambda_i^jh_j\, ,  \sum_{j\in \Lambda_1}\lambda_f^jh_j)\times K_{\mbox{\tiny{cl}}}(\sum_{j\in \Lambda_2}\lambda_i^jh_j\, ,  \sum_{j\in \Lambda_2}\lambda_f^jh_j)
\ee
much as in the clustering decomposition equation \eqref{equ:cluster}. 

 Even if the two regions don't decouple,  effective field theory (and universality) can in principle be used to write an effective average action at the scale $k$ \cite{Reuter} $\Gamma_k$, which ``integrates out" higher momentum modes and has an IR cutoff defined by the region we are studying. Using the effective action in place of the action, we can formally write:
\begin{align}
 K(\sum\lambda_i^jh_j,\sum\lambda_f^jh_j )&\approx K^{\lambda_j\leq\Lambda}_{\mbox{\tiny{cl}}}(\sum_{j\leq j_\Lambda}\lambda_i^jh_j\, ,  \sum_{j\leq j_\Lambda}\lambda_f^j h_j) \\ 
 &\times K^{\Lambda\leq\lambda_j\leq\Lambda'}_{\mbox{\tiny{cl}}}(\sum_{ j_\Lambda\leq \leq j_{\Lambda'}}\lambda_i^jh_j\, , \sum_{ j_\Lambda\leq j\leq j_{\Lambda'}}\lambda_f^jh_j)\times K^{\lambda_j\geq\Lambda'}_{\mbox{\tiny{cl}}}(\sum_{j\geq  j_\Lambda}\lambda_i^jh_j\, , \sum_{ j\geq j_\Lambda}\lambda_f^jh_j)\nonumber \end{align}
where e.g. $K^{\lambda_j\leq\Lambda}_{\mbox{\tiny{cl}}}$ uses the effective action at scale $k\sim \Lambda$. Such a factorization would also allow for cancellation of modes for which we have no direct access to in the ratio of amplitudes.\footnote{If no factorization is exact, i.e. if the higher momentum modes couple in a non-trivial way to the lower momentum ones, one could perhaps detect the presence of a fundamental high momentum cuttoff from interference effects of the lower momentum ones.} In any case, this is merely a tentative sketch of how the problem could be approached; these are issues that require much further study.

\section{Conclusions}
This paper is a companion paper to \cite{path_deco}, which did not deal with questions of locality.   The main point of this paper is that, freed from being a postulate, locality can now emerge from certain dynamical situations, and do so only approximately. 

%For example, the notion of kinematically locality does not  disallow space-time diffeomorphisms acting on the 4-metric $g_{\mu\nu}$. However, as explained in \cite{Wald_Lee}, there is no symplectic reduction of field space that represents those symmetries on the entire kinematic phase space. One can only project down the action of non-spatial space-time diffeomorphisms to phase space for the subspace of  \emph{4-dimensional field configurations that satisfy the equations of motion}.  It means that the symmetry  on phase space does not represent (off-shell) space-time refoliations, which from a quantum mechanical point of view would be required. Thus it opens the door for the approach advocated here (and elsewhere \cite{SD_first}), that refoliation invariance is also  regained only on-shell for solutions of the Hamiltonian dynamics, as an effective symmetry. One could always choose to have a strictly covariant space-time-based quantization procedure, in which case this incompatibility of the two pictures would of course not arise. However, in that case, it becomes difficult to deal with questions related to time-evolution of the fields themselves, since field space is given by the full space-time metric and one can only represent time evolution for a given background. This option -- of not pursuing a symplectic (dynamic) treatment of the space-time theory -- would be incompatible with my notion of dynamically generated locality.

I began by postulating configuration spaces which are formed by sections of tensor bundles. For a given region $O\subset M$, the subspace of field configurations which have a certain fixed boundary value, can be further algebraically decomposed into subspaces corresponding to the field configurations inside and outside $O$. This decomposition had to be done in a gauge-covariant manner.

That is because gauge degrees of freedom cloud matters of locality.\footnote{Note that it is difficult to tell whether an action is local or not by just looking at its Lagrangian. That is, it might be local in some gauge, while not in others. And looking at observables doesn\rq{}t really help, as in gravitational theories they are almost certainly non-local.}  I demanded that theories be what I termed \lq\lq{}kinematically local\rq\rq{}, meaning that the symmetries act as a group and not a groupoid, with their actions  depending solely on the local field configuration, not on the tangent bundle to configuration space. I.e. the action of the symmetries in this way should be local and could not depend on time derivatives of the fields and had to form a genuine group. This allowed us to form a principal fiber bundle out of configuration space. 

The split \eqref{equ:general_split} identifies physical regions ($O$ and $N$) with submanifolds in configuration space, $\mathcal{M}_O^{\partial O}, \mathcal{M}_N^{\partial O}$. These properties made it possible to form an equivalence relation by their action, and show that this equivalence relation obeyed the split \eqref{equ:general_split}, 
$$\frac{\mathcal{M}_{M}^{\partial O}}{\mathcal{G_M}}\simeq \frac{ \mathcal{M}_O^{\partial O}}{\mathcal{G}_O}\oplus \frac{ \mathcal{M}_N^{\partial O}}{\mathcal{G}_N}
$$

 Dynamics acts on configuration space and can thus  be non-local in physical space. But by looking at how it acts on the respective submanifolds in configuration space (pertaining to a region in physical space), e.g. $\mathcal{M}_O^{\partial O}$, it was possible to determine the circumstances under which it localizes those regions \emph{in physical space}, e.g. $O$. 
 This is the reason why we applied a dynamical criterion for which subsystems localize. 

%I have performed the study purely from a configuration space point of view, in which the global character of the system is manifest. But we can still recover structures in configuration space that represent the local degrees of freedom of a region in physical space. We do this by  identifying specific submanifolds in configuration space with equivalence classes of configurations. The equivalence class is made up of those configurations  which coincide in some region of physical space, thus establishing a link between regions in physical space and submanifolds in configuration space. 

   This led me to definition \ref{def:spatial_localization}.   Given extremal paths $\phi(t)$ over $\mathcal{M}_M^{\partial O}$, it is conditional on a comparison between their \emph{restriction}, $\mathsf{pr}_O\phi(t)$,  with \emph{intrinsic} extremal paths $\phi_{O}(t)$. Roughly,  the linear projection of the total dynamics to the local dynamics, according to \eqref{equ:general_split},  preserved the dynamics. It is a  condition analogous to:
$$e^{i\hat{H}t}\circ \mathsf{pr}_{\mbox{\tiny O}}= \mathsf{pr}_{\mbox{\tiny O}}\circ e^{i\hat{H}t}$$
 where $e^{i\hat{H}t}$ is the unitary time evolution operator,  the lhs denotes evolution for the intrinsic action in $O$, and the rhs denotes projection of the total evolution onto the $\mathcal{M}_O^{\partial O}$ subspace.    This ultimately led to an  analogy between, {\bf i}) a region of space being dynamically independent, and {\bf ii}) the submanifolds in configuration space related to this region being totally geodesic submanifolds (for systems whose Lagrangian is of Jacobi form).

This new understanding of locality is based solely on the properties of their extremal paths in configuration space. My ultimate aim was to \emph{not demand} locality from the start, as it is usually done,  but to have certain systems, under certain conditions,  exhibit it spontaneously. In this way non-locality becomes the norm for arbitrary systems, not the exception. \emph{We recover semi-classical local behavior when regions dynamically decouple from each other}. The definitions here also make it possible to study situations in which locality emerges for fundamentally non-local and/or non-relativistic systems, such as Horava-Lifschitz \cite{Horava}, Einstein-Aether \cite{Einstein_Aether},  shape dynamics \cite{SD_first}, etc.

There are three  important distinctions to notice in our present case, given in definition \ref{def:spatial_localization}: {\bf i)} it applies to more general actions, implicit, non-local,  and/or elliptic or parabolic in nature. Since the equations of motion are generically non-local in these cases, the definition delineates circumstances under which   this non-locality might be imperceptible to its local inhabitants, for particular field content.  {\bf ii)}  we can talk about an approximated sense in which the given system is local, by using the auxiliary supermetric on $\mathcal{M}_O^{\partial O}$. In other words, freed from being a postulate, locality can now emerge from certain dynamical situations, and do so only approximately.   {\bf iii)} It also makes it much more natural to treat cases in which multiple extremal histories (and thus quantum interference) occur, since  definition \ref{def:spatial_localization}  does not require a unique causal structure, and thus generalizes the usual relativistic concept, which does need a unique causal structure for the standard definition of domain of dependence. 
%A comparison with the more standard notion of domain of dependence for relativistic fields was made. The advantages of the present formulation manifest themselves in {\bf i}) the fact that using the inner product on configuration space one could have notions of ``approximate localization" --  not recoverable  by a theory where locality is postulated,  {\bf ii}) the localized region is explicitly defined by the \emph{dynamical}, as  opposed to kinematical,  field content, thus it can apply to theories that don\rq{}t have a universal light cone, or are not hyperbolic in character, and {\bf iii}) that it becomes much more natural to incorporate multiple extremal histories interpolating between a given initial and final field configuration (and thus quantum interference between partial field histories). 

I have furthermore shown that when two regions are \emph{mutually independent} the semi-classical path integral kernel decouples, proving cluster decomposition occurs in this case, which I take as a justification for the definitions themselves. This, one of the main results of this paper, is shown in proposition \ref{prop}.  It is important to note however, that the  regions must bilaterally  decouple for proposition \ref{prop} to hold.

%At some points of the text I have supposed  the action functional driving the dynamics to be put into Jacobi form. In this way there exists a metric in configuration space for which geodesics are exactly the extremal paths of the action. This made many analogies with concepts from Riemannian geometry possible, but it would be interesting to investigate which of these analogies can be replaced by more standard dynamical concepts from  symplectic geometry in the absence of the Jacobi metric.  

\section*{ACKOWLEDGMENTS}
I  would like to thank  Clement Delcamp for drawing the figures.  This research was supported  by Perimeter Institute for Theoretical Physics. Research at Perimeter Institute is supported by the Government of Canada through Industry Canada and by the Province of Ontario through the Ministry of Research and Innovation.

 \begin{appendix}
 \section*{APPENDIX}
 \subsection*{The Jacobi metric}\label{app:Jacobi_metric}

The Jacobi version of the Maupertuis principle establishes some instances when  dynamics
 can be viewed as geodesic motion in an associated Riemannian
manifold. That is, if the action can be written as
$$ S = \int (T - V)dt$$
for a system defined in a Riemannian manifold, $(M, g)$, with {\emph smooth} potential $U$, the extremal trajectories
of $S$ with energy $E = T + U$ coincide with the extremals (geodesics) of the
length functional 
$$L=\int ds
$$
 defined in $(M, h)$, where h is the Jacobi metric, conformally related to $g$ by $h = 2(E - V)g$.
This I will call the Jacobi procedure,  whereby one builds a metric in configuration space whose geodesics are extremal paths of the action.\footnote{ For classical non-relativistic particle mechanics, one could have that $V>E$, in which case one would obtain a purely imaginary action from the Jacobi procedure. This is related to the phenomenon of tunneling, which can be represented either as solutions of the Euclidean action (instantons) or from imaginary  valued solutions of the equations of motion \cite{Turok_real_time} in real time. See also \cite{Tanizaki} for a way to obtain tunneling phenomena from the real-time path integral.}
\subsection*{The semi-classical approximation}
Given an action $S$ in the configuration space of some field $\phi$ over a spatial manifold $M$ (we will discuss which fields we have in mind in the next sections), one can build the path integral propagator between an initial and final configuration, $\phi_1$ and $\phi_2$ as: 
\be\label{equ:path_integral}K[\phi_1,\phi_2]= \int_{\gamma\in\Gamma(\phi_1,\phi_2)} \mathcal{D}\gamma \exp{[iS[\gamma(\lambda)]/\hbar]}\ee
where $\Gamma(\phi_1,\phi_2)$ is taken to be the space of paths (in some differentiability class) between $\phi_1$ and $\phi_2$, and I used square brackets to denote functional dependence. 

The following semi-classical, or saddle point,  approximation  for the path integral in configuration space was the fundamental object used in \cite{path_deco}.  For (locally) extremal paths $\gamma_{\mbox{\tiny{cl}}}$ between an initial and a final field configuration, denoting the on-shell action for these paths as $S_{\gamma_{\mbox{\tiny{cl}}}}$, accurate for $1<< S_{\gamma_{\mbox{\tiny{cl}}}}/\hbar$, it  can be written as:  
\be\label{equ:semi_classical_exp} K_{{\mbox{\tiny{cl}}}}[{\phi^*}, \phi_f]= A \sum_{\gamma_{\mbox{\tiny{cl}}}}(\Delta_{\gamma_{\mbox{\tiny{cl}}}})^{1/2}\exp{\left(i S_{\gamma_{\mbox{\tiny{cl}}}}[{\phi^*},\phi_f]/\hbar\right)}
 \ee 
 where $A$ is a normalization factor (independent of the initial and final configurations),\footnote{In a finite-dimensional dynamical system of dimension $d$, this takes the form of the phase space volume occupied by a quantum state: $A=(2\pi\hbar)^{-d}$. } I used square brackets to denote functional dependence of the on-shell action,  and  the Van Vleck determinant is now defined as
 \be\label{equ:path_Van_vleck}
 \Delta_{\gamma_{\mbox{\tiny{cl}}}}:=\det\left(-\frac{\delta^2 S_{\gamma_{\mbox{\tiny{cl}}}}[{\phi^*},\phi_f]}{\delta {\phi^*}(x)\delta\phi_f(y)}\right)= \det\left(\frac{\delta\pi^\gamma_f[{\phi^*}; x)}{\delta {\phi^*}(y)}\right)
 \ee
where we used DeWitt's mixed functional/local  dependence notation $[{\phi^*}; x)$, and  the on-shell momenta is defined as 
$$\pi^\gamma_f[{\phi^*}; x):= -\frac{\delta S_{\gamma_{\mbox{\tiny{cl}}}}[{\phi^*},\phi_f]}{\delta \phi_f(x)}$$
The Van-Vleck matrix is degenerate when gauge-symmetries exist, in which case one must provide a suitable gauge-fixing.

\subsection*{Illustration of solution of Laplace equation on the annulus}
\begin{figure}[h]\label{figure:annulus}
\begin{center}
\includegraphics[width=10cm, height=6cm]{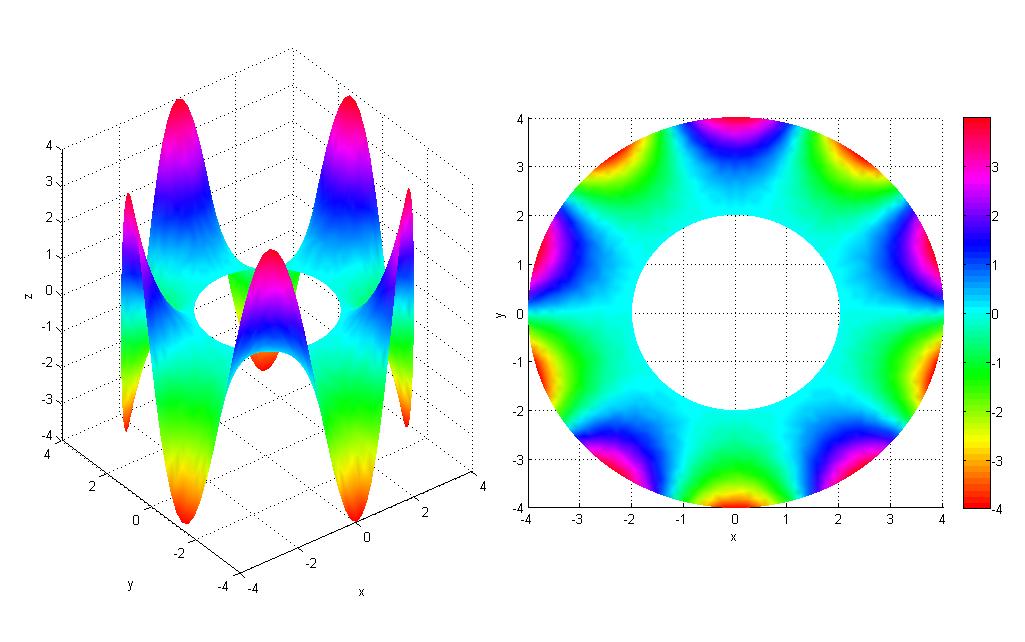}
\caption{Laplace\rq{}s equation on an annulus with radii $r_1=2, r_2=4$, with Dirichlet boundary conditions, $\phi(r_1)=0$ and  $\phi(r_2)=4\sin(5\theta)$.  Image credit: Wikimedia commons.}
\end{center}
\end{figure}

 \end{appendix}

%\bibliographystyle{ieeetr}

%\bibliography{decoherence}

\begin{thebibliography}{10}

\bibitem{path_deco}
H.~Gomes, ``{Path integrals in configuration space and the emergence of
  classical behavior for closed systems},'' 2015.

\bibitem{conformal_geodesic}
H.~Gomes, ``A geodesic model in conformal superspace,'' {\em To appear}, 2016.

\bibitem{Isham_parametrized}
C.~Isham and K.~Kuchar, ``Representations of spacetime diffeomorphisms. ii.
  canonical geometrodynamics,'' {\em Annals of Physics}, vol.~164, no.~2,
  pp.~316 -- 333, 1985.

\bibitem{Michorbook}
A.~Kriegl and P.~W. Michor, {\em The Convenient Setting of Global Analysis}.
\newblock American Mathematical Society, Providence, 1997.

\bibitem{Freed}
D.~S. Freed and D.~Groisser, ``{The basic geometry of the manifold of
  Riemannian metrics and of its quotient by the diffeomorphism group.},'' {\em
  Mich. Math. J.}, vol.~36, no.~3, pp.~323--344, 1989.

\bibitem{Fischer}
A.~R. Fischer, ``The theory of superspace,'' in {\em Proceedings of the
  Relativity Conference held 2-6 June, 1969 in Cincinnati, OH. Edited by Moshe
  Carmeli, Stuart I. Fickler, and Louis Witten. New York: Plenum Press, 1970.,
  p.303}, 1970.

\bibitem{Giulini}
D.~Giulini, ``{What is the geometry of superspace?},'' {\em Phys. Rev.},
  vol.~D51, pp.~5630--5635, 1995.

\bibitem{Ebin}
D.~Ebin, ``The manifold of riemmanian metrics,'' {\em Symp. Pure Math., AMS,},
  vol.~11,15, 1970.

\bibitem{Palais}
R.~Palais, ``On the existence of slices for the actions of non-compact
  groups,'' {\em Ann. of Math.}, vol.~73, pp.~295--322, 1961.

\bibitem{Trudinger}
D.~Gilbarg and N.~Trudinger, {\em Elliptic Partial Differential Equations of
  Second Order}.
\newblock Springer, 2001.

\bibitem{Gil-Medrano}
O.~Gil-Medrano and P.~W. Michor, ``The riemannian manifold of all riemannian
  metrics,'' {\em Quarterly Journal of Mathematics, (42), 183-202}, 1991.

\bibitem{Wald_Lee}
J.~Lee and R.~M. Wald, ``Local symmetries and constraints,'' {\em Journal of
  Mathematical Physics}, vol.~31, no.~3, pp.~725--743, 1990.

\bibitem{ADM}
R.~Arnowitt, S.~Deser, and C.~Misner, ``The dynamics of general relativity
  pp.227�264,,'' in {\em in Gravitation: an introduction to current research,
  L. Witten, ed.}, Wiley, New York, 1962.

\bibitem{SD_first}
H.~Gomes, S.~Gryb, and T.~Koslowski, ``{Einstein gravity as a 3D conformally
  invariant theory},'' {\em Class. Quant. Grav.}, vol.~28, p.~045005, 2011.

\bibitem{Schroedinger}
E.~Schroedinger, ``Die gegenwartige situation in der quantenmechanik,'' {\em
  Naturwissenschaften}, vol.~23, no.~48, pp.~807--812, 1935.

\bibitem{RT}
T.~Regge and C.~Teitelboim, ````role of surface integrals in the hamiltonian
  formulation of general relativity",'' {\em Annals of Physics}, vol.~{\bf 88},
  1974.

\bibitem{Reuter}
M.~Reuter, ``{Nonperturbative evolution equation for quantum gravity},'' {\em
  Phys.Rev.}, vol.~D57, pp.~971--985, 1998.

\bibitem{geodesic_foliation}
D.~Johnson and L.~Whitt, ``Tottaly geodesic foliations,'' {\em J. of
  Differential Geometry.}, vol.~15, pp.~225--235, 1980.

\bibitem{Horava}
P.~Horava, ``{Quantum Gravity at a Lifshitz Point},'' {\em Phys.Rev.},
  vol.~D79, p.~084008, 2009.

\bibitem{Einstein_Aether}
C.~Eling, T.~Jacobson, and D.~Mattingly, ``{Einstein-Aether theory},''
  pp.~163--179, 2004.

\bibitem{Turok_real_time}
N.~Turok, ``{On Quantum Tunneling in Real Time},'' {\em New J. Phys.}, vol.~16,
  p.~063006, 2014.

\bibitem{Tanizaki}
Y.~Tanizaki and T.~Koike, ``Real-time feynman path integral with picard
  lefschetz theory and its applications to quantum tunneling,'' {\em Annals of
  Physics}, vol.~351, pp.~250 -- 274, 2014.

\end{thebibliography}

\end{document}